# Negative refraction index of the quantum lossy left-handed transmission lines affected by the displaced squeezed Fock state and dissipation

Shun-Cai Zhao,[1,][*] Xiao-Jing Wei,[1] and Qi-Xuan Wu[2]

[1]*Department of Physics, Faculty of Science,*
*Kunming University of Science and Technology, Kunming, 650500, PR China*
[2]*Faculty of Foreign Languages and culture,*
*Kunming University of Science and Technology, Kunming, 650500, PR China*

(Dated: June 23, 2017)

## Abstract

Quantum lossy left-handed transmission lines (LHTLs) are central to the miniaturized application in microwave band. This work discusses the NRI of the quantized lossy LHTLs in the presence of the resistance and the conductance in a displaced squeezed Fock state (DSFS). And the results show some novel specific quantum characteristics of NRI caused by the DSFS and dissipation, which may be significant for its miniaturized application in a suit of novel microwave devices.

---

[*] Corresponding author: zhaosc@kmust.edu.cn; zscnum1@126.com

In one dimension, left-handedness is described as the double opposite orientations between the wave vector $\vec{k}$ and Poynting vector, the phase and group velocity corresponding to the dispersion relation $\partial\omega(k)/\partial k < 0$[1]. The left-handedness in the LHTL can be achieved by a discrete array of series capacitors and parallel inductors[2–4](Fig.1), which remains the metamaterial via the array of series dual interchanged inductors and capacitors other than the usual discrete representation of the right-handed transmission line (RHTL). LHTL is perhaps one of the most representative and potential candidates due to its non-resonant type with low loss, broad operating frequency band, as well as planar configuration[5, 6], which is often related with easy fabrication for NRI applications in a suite of novel guided-wave[7], radiated-wave[8], and refracted-wave devices and structures[9, 10].

Parallel to these developments are those of technological miniaturization of circuits, and their mesoscopic aspects become more relevant. Especially, the scale of fabricated circuits reached to Fermi wavelength, quantum mechanical properties [11, 12] become important while the application of classical mechanics fails. And one of the quantum theories for these mesoscopic circuits was proposed by Li and Chen[13] where charge discreteness was considered explicitly, which is different to the quantization scheme from Louisell [14]. Louisell didn't quantize the charges and currents while suggested an ideal one-dimensional transmission line that can be described by a classical wave equation and then can be quantized by the standard canonical quantization approach.

In our previous work[15], we discussed the NRI of the quantized ideal lossless LHTL dependent the fluctuations of the current, photons, frequency and temperature via the thermal field dynamics (TFD) theory[16, 17]. The results show that the NRI can be facilitated when the travelling electromagnetic wave operates at a lower frequency and temperature with less photons in microwave frequency band. In this article, the quantum properties of NRI in the lossy transmission line is discussed in a displaced squeezed Fock state(DSFS). The DSFS can be produced by the applications of optical and electric fields in some physical system[18–21] which can exhibit squeezed Fock state, displaced Fock state, squeezed vacuum state, coherent state, squeezed state, Fock state and vacuum state[22] and many nonclassical effects are revealed in these states[23–26]. The quantum effect on the NRI of the quantum lossy LHTLs will be shown in the DSFS mathematically.

The paper is organized as follows. In Sec. 2, we quantize the traveling wave in the unit cell circuit of the lossy LHTLs. The NRI of the lossy LHTLs is deduced in the DSFE in Sec. 3, and we evaluate the NRI dependent the squeezed parameters and dissipative parameters in Sec. 4. Section 5 presents our summary and conclusions.

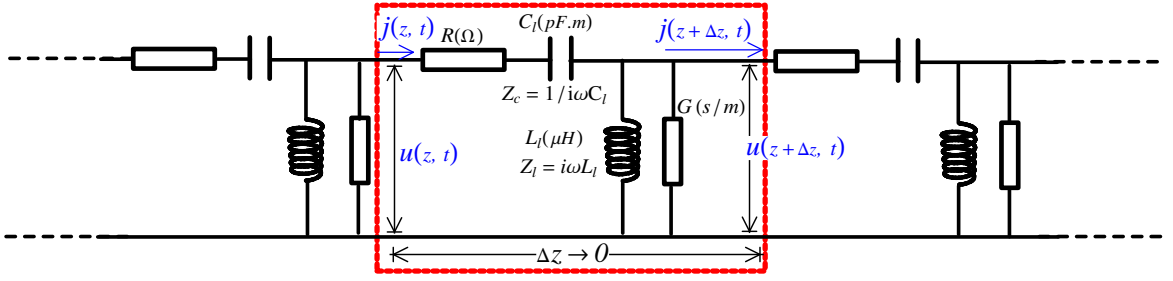

FIG. 1. (Color online) Schematic diagram of equivalent circuit for the dissipative LHTL. The red dashed block corresponds to the unit cell circuit for the LHTL.

## I. QUANTIZATION OF TRAVELING WAVE IN THE LHTLS

As shown in Fig.1, the one dimensional homogeneous LHTLs are composed by the array of series interchanged roles of inductors and capacitors other than the usual RHTL. The diagram in the red dashed block shows the unit cell circuit($\Delta z \to 0$) of this LHTL. Comparing with our previous work[15], the imported resistance $R$ and conductivity $G$ represent the dissipation in each unit cell circuit. $C_l$ ($pF\cdot$ m) and $L_l$ ($\mu H\cdot$ m) are the capacitance and inductance of a unit cell of the LHTL. So the impedances and admittances in per unit length corresponding to the capacitance and inductance are $Z = 1/i\omega C_l$ and $Y = 1/i\omega L_l$, respectively. According to the Fig.1, the Kirchhoff's law bringing out the classical motion equations in per unit length are as follows,

$$\frac{\partial u(z,t)}{\partial z} = -j(z,t)[R + \frac{1}{i\omega C_l}] \tag{1}$$

$$\frac{\partial j(z,t)}{\partial z} = -u(z,t)[G + \frac{1}{i\omega L_l}] \tag{2}$$

where $u(z,t)$ is the voltage, $j(z,t)$ is the current, and $\omega$ is the angle frequency. Considering the separated current between the inductance $L_l$ and conductivity $G$, the voltage of the inductance $L_l$ can be reached to $u(z,t) = \frac{L_l}{\omega L_l G + 1}\frac{\partial j(z,t)}{\partial t}$. Eliminating the current or voltage in the above two equations to decouple in the usual way, we obtain the forward plane-wave solutions to current and voltage in the following,

$$j(z,t) = \exp{(-\alpha z)}[\mathbb{A}\exp{(i\beta z - i\omega t)} + \mathbb{A}^*\exp{(i\omega t - i\beta z)}], \tag{3}$$

$$u(z,t) = \frac{i\omega L_l}{\omega L_l G + 1}\exp{(-\alpha z)}[\mathbb{A}^*\exp{(i\omega t - i\beta z)} - \mathbb{A}\exp{(i\beta z - i\omega t)}] \tag{4}$$

in which $\sqrt{(R + \frac{1}{i\omega C_l})(G + \frac{1}{i\omega L_l})} = \alpha + i\beta$. In Eq.(3) and Eq.(4), $\mathbb{A}^*$ is the complex conjugate of $\mathbb{A}$ for normalization purposes. To simplify Eq.(3) and Eq.(4), two functions $\eta(z,t)$ and $\zeta(z,t)$ are introduced by

$$j(z,t) = \exp(-\alpha z)\eta(z,t),$$
$$u(z,t) = \frac{\omega^2}{z_0}\exp(-\alpha z)\zeta(z,t)$$

where $z_0$ is the length of per unit cell circuit. We assume that $\omega$ is given and choose the unit length of circuit $z_0$ to be a fixed integral number of wavelengths $z_0 = m\lambda$. Then, $\eta(z,t)$ and $\zeta(z,t)$ differential $t$ are

$$\frac{\partial \eta(z,t)}{\partial t} = \frac{\omega^2(\omega L_l G + 1)}{L_l z_0}\zeta(z,t), \tag{5}$$

$$\frac{\partial \zeta(z,t)}{\partial t} = -\frac{L_l z_0}{(\omega L_l G + 1)}\eta(z,t) \tag{6}$$

It is straightforward to find the following relation,

$$\frac{\partial}{\partial \eta(z,t)}\left(\frac{\partial \eta(z,t)}{\partial t}\right) + \frac{\partial}{\partial \zeta(z,t)}\left(\frac{\partial \zeta(z,t)}{\partial t}\right) = 0$$

So, $\eta(z,t)$ and $\zeta(z,t)$ are the canonically conjugate variables and obey the hamiltonian canonical equations,

$$\frac{\partial \eta}{\partial t} = \frac{\partial H}{\partial \zeta}, \tag{7}$$

$$\frac{\partial \zeta}{\partial t} = -\frac{\partial H}{\partial \eta} \tag{8}$$

The integrals of Eq.(7) and Eq.(8) can deduce the hamiltonian for the electromagnetic wave with the ignored integral initial constant.

$$H = \frac{\omega^2(\omega L_l G + 1)}{2L_l z_0}\zeta^2 + \frac{L_l z_0}{2(\omega L_l G + 1)}\eta^2$$

If we associate hermitian operators of $\hat{\eta}(z,t)$ and $\hat{\zeta}(z,t)$ and require that they satisfy the commutation relation,

$$[\hat{\eta}, \hat{\zeta}] = \frac{2iL_l z_0}{\omega(\omega L_l G + 1)}[\hat{\mathbb{A}}, \hat{\mathbb{A}}^*] = i\hbar \tag{9}$$

and we achieve the quantized lossy unit cell circuit with the definitions,

$$\hat{\mathbb{A}} = \hat{a}\sqrt{\frac{\hbar\omega(\omega L_l G + 1)}{2L_l z_0}},$$

$$\hat{\mathbb{A}^*} = \hat{a^*}\sqrt{\frac{\hbar\omega(\omega L_l G + 1)}{2L_l z_0}}$$

Substituting the above equations into Eq.(9), the quantum condition $[\hat{a}, \hat{a^*}] = 1$ is obtained. Then the two operators $\hat{\eta}$ and $\hat{\zeta}$ are,

$$\hat{\eta}(z,t) = \sqrt{\frac{\hbar\omega(\omega L_l G + 1)}{2L z_0}}[\hat{a}\exp(i\beta z - i\omega t) + \hat{a^\dagger}\exp(i\omega t - i\beta z)], \tag{10}$$

$$\hat{\zeta}(z,t) = i\sqrt{\frac{\hbar L_l z_0}{2\omega(\omega L_l G + 1)}}[\hat{a^\dagger}\exp(i\omega t - i\beta z) - \hat{a}\exp(i\beta z - i\omega t)] \tag{11}$$

and the quantum Hamiltonian of the unit cell circui can be written as $\hat{H} = \hbar\omega(\hat{a^\dagger}\hat{a} + \frac{1}{2})$ which is analogous to the oscillator's Hamiltonian operator in Schrödinger representation.

## II. NRI IN THE DSFS

We give the DSFS $|z, \xi, n\rangle$ defined by $|z, \xi, n\rangle = \hat{D}(z)\hat{S}(\xi)|n\rangle$[21, 25], where the operator $\hat{D}(z) = \exp(z\hat{a}^\dagger - z^*\hat{a})$ is the displacement operator, $z = |z|exp(i\theta)$ ($|z| > 0, 0 \leq \theta < 2\pi$) is the displacement parameter. $\hat{S}(\xi) = \exp(\frac{1}{2}\xi\hat{a^\dagger}^2 - \frac{1}{2}\xi^*\hat{a}^2)$ is the squeeze operator, $\xi = |\xi|exp(i\phi)$ ($|\xi| > 0, 0 \leq \phi < 2\pi$) is known as the squeezed parameter, and $\phi$ indicates the direction of the squeezing. In which $\hat{a}$ ($\hat{a}^\dagger$) is the annihilation (creation) operator of the boson field.

Using the formula ( $e^{\lambda\hat{A}}\hat{B}e^{-\lambda\hat{A}} = \hat{B} + \lambda[\hat{A},\hat{B}] + \frac{\lambda^2}{2}[\hat{A},[\hat{A},\hat{B}]] + \cdots$), it's easily to prove the following relations,

$$\hat{D}^\dagger(z)\hat{a}\hat{D}(z) = \hat{a} + z,$$
$$\hat{D}^\dagger(z)\hat{a^\dagger}\hat{D}(z) = \hat{a^\dagger} + z^*,$$
$$\hat{S}^\dagger(\xi)\hat{a}\hat{S}(\xi) = \hat{a}\cosh|\xi| + \hat{a^\dagger}e^{i\phi}\sinh|\xi|,$$
$$\hat{S}^\dagger(\xi)\hat{a^\dagger}\hat{S}(\xi) = \hat{a^\dagger}\cosh|\xi| + \hat{a}e^{-i\phi}\sinh|\xi|$$

Therefore, the mean and mean-square values of current in DSFS in Heisenberg picture can be obtained, respectively,

$$\langle \hat{j} \rangle = M e^{-\alpha z}[|z|e^{i\theta}e^{(i\beta z)} + |z|e^{-i\theta}e^{(-i\beta z)}], \quad (12)$$

$$\langle \hat{j}^2 \rangle = F e^{-2\alpha z}[[(2n+1)\cosh|\xi|\sinh|\xi|e^{i\phi} + z^2]e^{2i(\beta z)} + [(2n+1)\cosh|\xi|\sinh|\xi|e^{-i\phi}$$
$$+ z^{*2}]e^{2i(-\beta z)} + 2(n+1)\cosh^2|\xi| + 2n\sinh^2|\xi| + 2zz^*] \quad (13)$$

where $M = \sqrt{\frac{\hbar\omega(\omega L_l G+1)}{2L_l z_0}}$, and $F = \frac{\hbar\omega(\omega L_l G+1)}{2L_l z_0}$. Then the quantum fluctuation of current of the unit cell circuit of the LHTL in the DSFS reaches to,

$$\langle (\Delta \hat{j})^2 \rangle = F e^{-2\alpha z}[(2n+1)\frac{1}{2}\sinh|2\xi|e^{i\phi}e^{2i\beta z} +$$
$$(2n+1)\frac{1}{2}\sinh|2\xi|e^{-i\phi}e^{-2i\beta z}$$
$$+ (2n+1)\cosh|2\xi| + 1] \quad (14)$$

Similarly, we can deduce the quantum fluctuation of voltage but we ignore it here. It's notice that $\beta z$ is infinitesimal when $z$ is a dimensionless, and the relation ($N = \frac{c_0 \beta}{\omega}$[28], $c_0$ is the light speed in vacuum) between propagation constant and refraction index in LHTL. The refraction index can be deduced by Eq.(14),

$$N = -\frac{c_0}{2\omega z \sin\phi}[\frac{2L_l z_0 e^{2\alpha z} - \hbar\omega^2 L_l G}{\hbar\omega(2n+1)(\omega L_l G+1)\sinh|2\xi|}\langle(\Delta\hat{j})^2\rangle - \frac{\cosh|2\xi|}{\sinh|2\xi|} - \cos\phi] \quad (15)$$

## III. NUMERICAL RESULTS AND DISCUSSION

In the common sense, the NRI of the classic LHTL circuit should remain a steady value in its equilibrium state. While the unit cell circuit of the LHTL approaches to the mesoscopic dimension, the quantum fluctuation may change the characteristic [11, 12]. And this can be verified by the NRI expression in the Eq.(15). In the following, we'll explore the dependence of the NRI on DSFS and dissipative parameters.

### A. NRI squeezed by the displacement and direction parameters

In our model, the mesoscopic dimension of the unit cell circuit is a prominent feature of the quantum lossy LHTL. The distribution properties of NRI along the transmission line is an attractive characteristic. Because the DSFS can provide plenty of quantum states, many mesoscopic systems were investigated in it theoretically and experimentally [18–21, 23–26]. Fig.2 shows the evolution of NRI in DSFS (n=5) and displaced squeezed state (n=0) within the length of unit cell circuit $z_0$.

It notes that NRI isn't homogeneous with different squeezing parameters along the unit length in Fig.2. In the range of [0, 0.2$z_0$], NRI declines sharply while it is asymptotic to a steady value in the following ranges in both DSFS (n=5) and displaced squeezed state (n=0). Meanwhile, the squeezing parameters play an important role and their squeezing influences are obvious: when the values of the squeezing parameters increases by $0.05\pi$, the values of NRI are squeezed greatly in DSFS (n=5). And the squeezing parameters cast a similar characteristic on the NRI in the displaced squeezed state (n=0). However, we also notice a different feature of NRI in the displaced squeezed state: the values of NRI are much small than its counterpart of the DSFS (n=5). This may be illustrated by the quantum Hamiltonian $\hat{H} = \hbar\omega(\hat{n}+\frac{1}{2})$. The little photon numbers means a weak electromagnetic wave field, the weaker field leads to the smaller NRI. The NRI alone the unit cell circuit exhibits the counter-intuitive features, which are different from its classic counterpart.

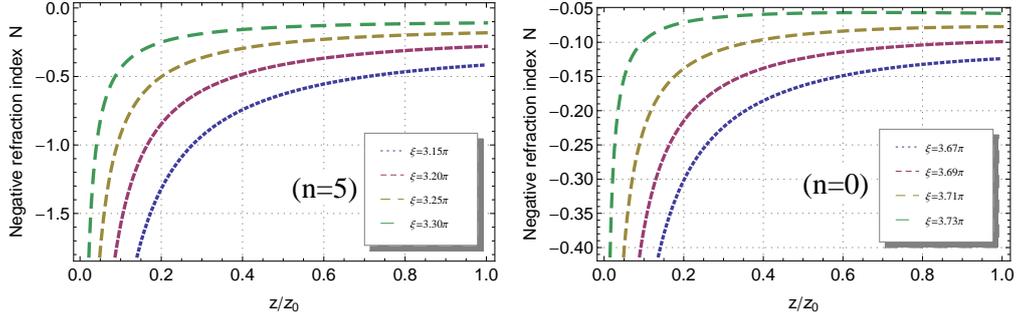

FIG. 2. (Color online) NRI dependent the lengths of the circuit $z/z_0$ via different squeezing parameters $\xi=3.15\pi, 3.20\pi, 3.25\pi, 3.30\pi$ in DSFS (n=5) and $\xi=3.67\pi, 3.69\pi, 3.71\pi, 3.73\pi$ in the displaced squeezed state (n=0), respectively. The other parameters are $C_l = 398pF$, $L_l = 995\mu H$, $R = 0.1\Omega$, $G = 0.5 \times 10^{-2} S/m$, $z_0 = 4\mu m$, $\phi = \pi/3$, $\langle(\Delta\hat{j})^2\rangle = 1$, $\omega = 2.5GHz$.

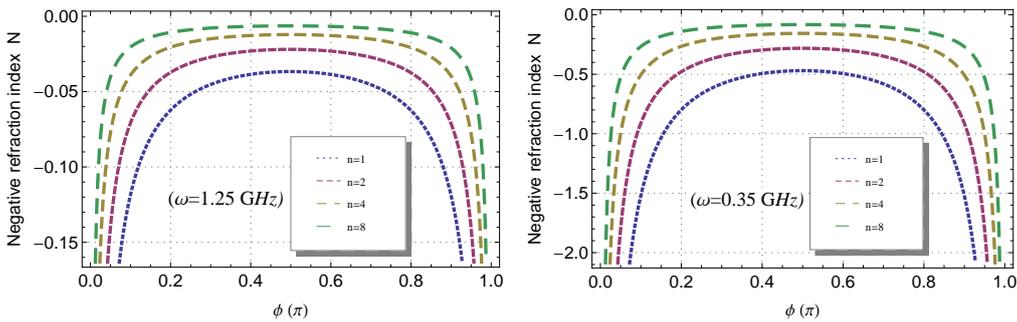

FIG. 3. (Color online) NRI dependent squeezing directions $\phi$ via different photons numbers $n=1, 2, 4, 8$ in DSFS with $\omega=1.25GHz$, $0.35GHz$, respectively. $\xi=2.80\pi$, $R = 0.2\Omega$, and the other parameters are same as in Fig.2.

Another interesting phenomena is the squeezing direction in the DSFS, which is a purely quantum behavior. Fig.3 shows the effect of the squeezing direction on NRI at $\omega=1.25GHz$ and $0.35GHz$, respectively. The two gray blocks correspond to two included angle intervals $[0, 0.2\pi]$ and $[0.8\pi, 1.0\pi]$ alone the transmission line, respectively, which indicates two opposite squeezed directions. The interval of $[0, 0.2\pi]$ means the squeezed direction along transmission line, and a reverse squeeze happens in the interval of $[0.8\pi, 1.0\pi]$. It notes that NRI decreases sharply when the squeezing direction along the transmission line($\phi\epsilon[0, 0.2\pi]$), while the NRI booms in the squeezing reverse direction($\phi\epsilon[0.8\pi, \pi]$). And the photon numbers play a reverse linear role on the NRI, i.e., the larger quantity of photon brings out the lesser NRI. The result is distinctive brought by the squeezing parameters $\xi$ on the NRI.

## B. NRI affected by the resistance and frequency

In the following, we will show the circuit's parameters, i.e., the resistance $R$ effect on the NRI. In generally, the resistance $R$ is not a desired role in electricity. Out of the resistance $R$ brings out Joule heat via the Ohm Law. The presence of the resistance and the conductance represents the dissipation in our quantum LHTL model, and their relation is set as $G = \frac{1}{R} \times 10^{-2}$. As shown in Fig.4, it notes that there is a booming range $[0, 0.5\Omega]$ (See the gray blocks ) and a flat slope $[0.5\Omega, 3\Omega]$ for NRI both in DSFS (N=5) and in displaced squeezed state (n=0) via the different squeezing parameters. In the gray blocks of Fig.4, the resistance $R$ brings out the increasing NRI. What's more, we notice that the squeezing effect is still prominent. The intense squeezing effect can lead to a small NRI. In the quantized unit cell circuit of the LHTL, the resistance violates the classic Ohm Law and produces a novel performances for NRI.

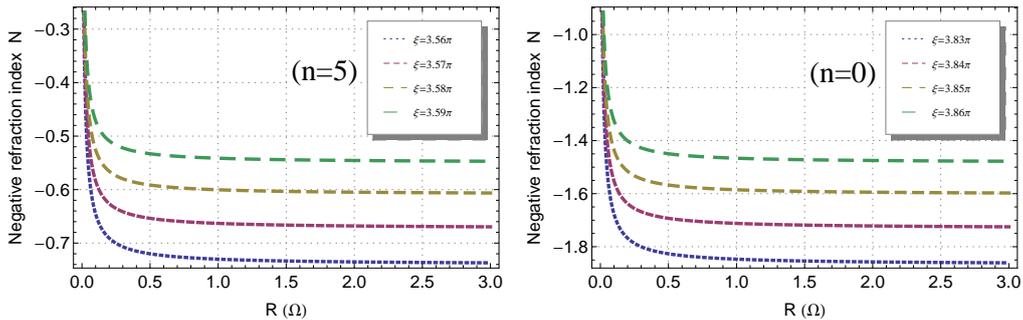

FIG. 4. (Color online) NRI dependent the resistances $R(\Omega)$ via different squeezing parameters $\xi=3.56\pi, 3.57\pi, 3.58\pi, 3.59\pi$ in DSFS (n=5) and $\xi=3.83\pi, 3.84\pi, 3.85\pi, 3.86\pi$ in the displaced squeezed state (n=0), respectively. $\langle(\Delta\hat{j})^2\rangle = 10$, $z = 1\mu m$, $G = \frac{1}{R} \times 10^{-2}$, and the other parameters are same as in Fig.2.

Another novel feature is the influence of frequency on NRI. The frequency of the electromagnetic wave travelling in the LHTL is an important measure indicator. As mentioned in the previous work[9, 28], the left-handedness appears only in the microwave frequency band. That the left-handedness in the microwave frequency band is homogeneous or not is not clear in the quantized unit cell circuit of the LHTL. Fig.5 illustrates that the NRI decreases sharply in the frequency band of [0, 1 GHz], while it reaches to the stable value in the range of [1 GHz, 2 GHz] under the two different squeezed parameters $\xi=3.0\pi$, $3.2\pi$, respectively. The NRI obtains the larger values in the low frequency bands while the smaller values in the increasing frequencies. And the value ranges of NRI are [0, -2.5] and [0, -0.4] with the squeezed parameters $\xi=3.0\pi$, $3.2\pi$, respectively. The squeezing effect on NRI is still prominent in Fig.5.

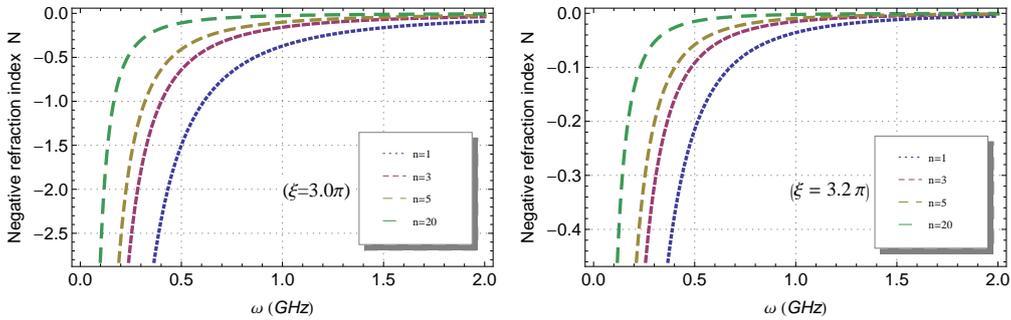

FIG. 5. (Color online) NRI dependent the frequency $\omega(GHz)$ via different photon numbers $n=1, 3, 5, 20$ in DSFS with $\xi=3.0\pi$, $3.2\pi$, respectively. $z = 2\mu m$, $R = \frac{1}{G}= 0.2\Omega$, and the other parameters are same as in Fig.4.

To conclude this article let us recall the squeezing effect and the dissipation in the quantum unit cell circuit of the LHTL. The increasing squeezing effect can lead the decreasing NRI. While the dissipative factor, i.e., the resistance or conductance exhibits specific quantum characteristic. What's more, the frequency of the electromagnetic travelling wave and its energy (i.e., the photon numbers) also play their specific quantum roles in the quantum LHTL. These novel quantum effects deserve the further research on miniaturization application of LHTL.

## IV. CONCLUSION

In this paper, we focused on the quantum characteristic of NRI dependent the squeezing effect and the dissipation in the unit cell circuit of the quantum LHTL. A dramatic squeezing effect is shown both by the displaced parameters and by the squeezing direction. And the NRI exhibits an increasing dependence within a small scale ranges, and a asymptotic steady dependence in the following value intervals of the resistance, while the NRI presents the non-homogeneous quantum distribution in the microwave frequency bands. These revealed novel quantum effects on NRI should be paid much attention in the further research on the miniaturization application of LHTL.

ACKNOWLEDGMENTS

This work is supported by the National Natural Science Foundation of China ( Grant Nos. 61205205 and 6156508508 ), the General Program of Yunnan Applied Basic Research Project, China ( Grant No. 2016FB009 ) and the Foundation for Personnel training projects of Yunnan Province, China ( Grant No. KKSY201207068 ).